\documentclass[9pt,twocolumn,twoside]{osajnl}

\journal{ol} 

\setboolean{shortarticle}{true}

\title{Chiral condensates in a polariton hexagonal ring}

\author[1,*]{Xuekai Ma}
\author[2,3]{Yaroslav V. Kartashov}
\author[4,5,6]{Alexey Kavokin}
\author[1,7]{Stefan Schumacher}

\affil[1]{Department of Physics and Center for Optoelectronics and Photonics Paderborn (CeOPP), Universit\"{a}t Paderborn, Warburger Strasse 100, 33098 Paderborn, Germany}
\affil[2]{Institute of Spectroscopy, Russian Academy of Sciences, Troitsk, Moscow, 108840, Russia}
\affil[3]{Russian Quantum Center, Skolkovo 143025, Russia}
\affil[4]{Westlake University, School of Science, 18 Shilongshan Road, Hangzhou 310024, Zhejiang Province, China}
\affil[5]{Westlake Institute for Advanced Study, Institute of Natural Sciences,
18 Shilongshan Road, Hangzhou 310024, Zhejiang Province, China}
\affil[6]{Spin Optics Laboratory, St-Petersburg State University, 1, Ulianovskaya, 198504 St-Petersburg, Russia}
\affil[7]{College of Optical Sciences, University of Arizona, Tucson, AZ 85721, USA}
\affil[*]{Corresponding author: xuekai.ma@gmail.com}

\begin{abstract}
We model generation of vortex modes in exciton-polariton condensates in semiconductor micropillars, arranged into a hexagonal ring molecule, in the presence of TE-TM splitting. This splitting lifts the degeneracy of azimuthally modulated vortex modes with opposite topological charges supported by this structure, so that a number of non-degenerate vortex states characterized by different combinations of topological charges in two polarization components appears. We present a full bifurcation picture for such vortex modes and show that because they have different energies, they can be selectively excited by coherent pump beams with specific frequencies and spatial configurations. At high pumping intensity, polariton-polariton interactions give rise to the coupling of different vortex resonances and a bistable regime is achieved.
\end{abstract}
\setboolean{displaycopyright}{true}

\begin{document}

\maketitle

Microcavity polaritons are bosonic quasiparticles with a finite lifetime on a picosecond scale. They may condense~\cite{deng2002condensation,kasprzak2006bose}, but still experience spontaneous decay accompanied by the emission of coherent light, the phenomenon called polariton lasing~\cite{PhysRevA.53.4250}. Especially interesting is the situation where condensation and lasing occur in states carrying nonzero topological charges. In this case, spatial structuring of the microcavity potential energy landscape, the presence of spin-orbit interaction (SOI), and strong polariton-polariton interactions may dramatically affect the emerging polariton states.

SOIs of different physical origin have been widely studied and play a crucial role in many areas of physics, including physics of semiconductors~\cite{wolf2001spintronics} and optics~\cite{bliokh2015spin,zambon2019optically}. It can also strongly affect the behaviour of excitations in optoelectronic systems, such as semiconductor microcavities operating in the regime of strong coupling between quantum-well excitons and cavity photons, where exciton-polariton condensates are formed~\cite{PhysRevB.59.5082,PhysRevLett.92.017401,PhysRevX.5.011034}. The SOI for polaritons arises from the splitting of the transverse-electric (TE) and transverse-magnetic (TM) modes of the cavity photons. Many interesting phenomena caused by the TE-TM splitting were reported for microcavity polaritons, such as the formation of half-quantum vortices~\cite{PhysRevLett.99.106401,lagoudakis2009observation,PhysRevB.94.134310,PhysRevB.101.205301}, optical spin Hall effect~\cite{PhysRevLett.95.136601,leyder2007observation}, skyrmions~\cite{PhysRevLett.110.016404,donati2016twist}, and topological insulators~\cite{PhysRevLett.114.116401, PhysRevB.91.161413,PhysRevX.5.031001,kartashov2016modulational,klembt2018exciton,PhysRevLett.122.083902,zhang2019finite}.
The resonant excitation of states with different topological charges for different spin components of a polariton condensate has been proposed~\cite{PhysRevB.75.241301}. In this context, it is important to develop the tools for topological engineering of polariton condensates that would enable on demand generation of vortices with specific topological charges~\cite{ma2020realization}.

In this Letter, we study the formation of vortex polariton modes in micropillars arranged into a hexagonal structure (molecule) in the presence of TE-TM splitting. Vortex modes with opposite topological charges in such a potential in the absence of the TE-TM splitting are degenerate, i.e. they have identical energies. The TE-TM splitting lifts the degeneracy affecting energies of states in different ways: while several vortices with allowed topological charges acquire increased frequencies, most of them are pushed to lower frequencies. This enables excitation of desired vortex states by coherent pumping with trivial phase distribution. The excitation efficiency depends not only on the frequency of the coherent pump, but also on its spatial configuration and symmetry. We also investigate the influence of the nonlinearity caused by polariton-polariton interactions on the modes and demonstrate the bistability regime.

\begin{figure*} [t]
\centering
{\includegraphics[width=1\linewidth]{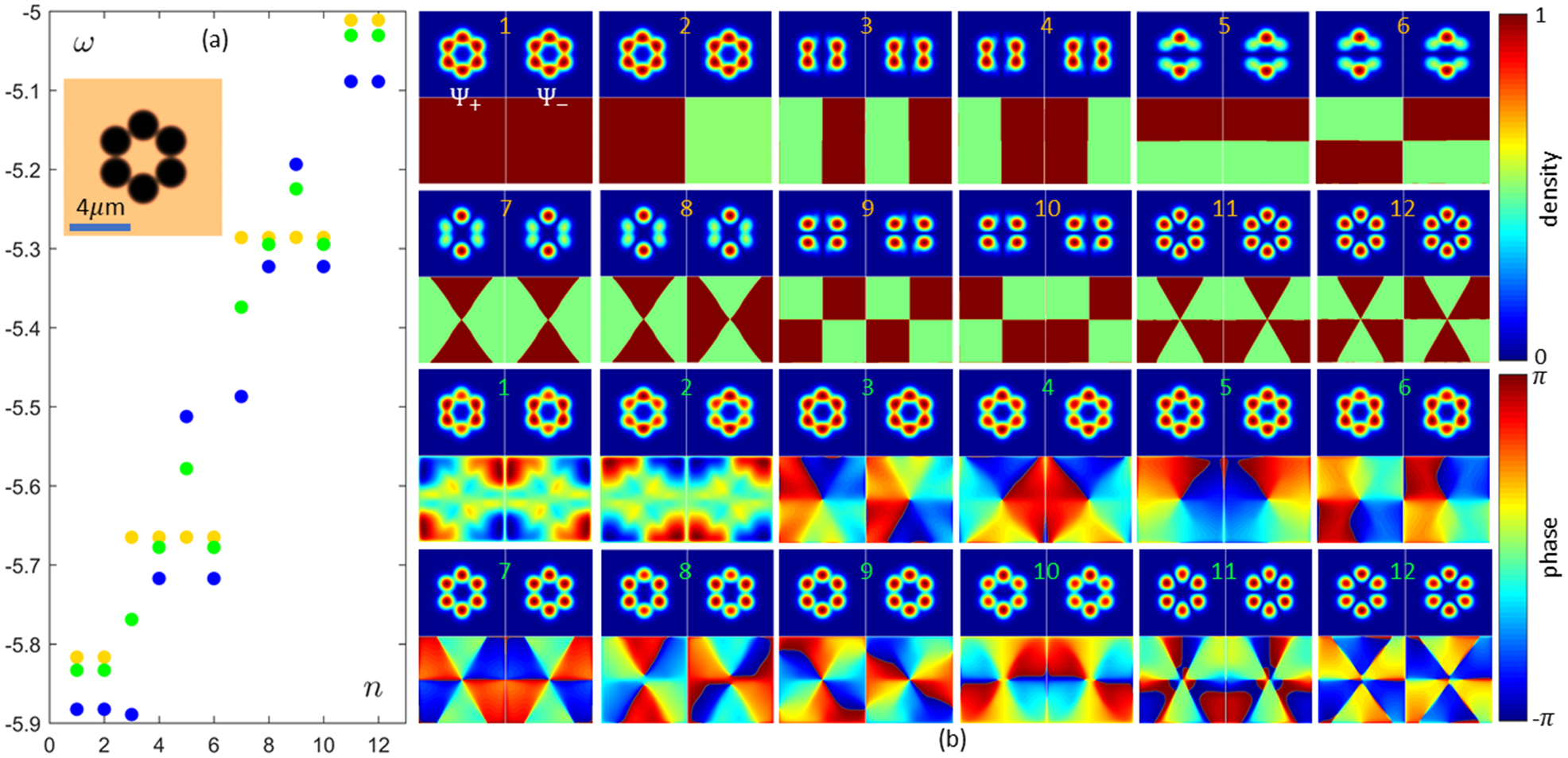}}
\caption{\textbf{Eigenstates of polaritons in a hexagonal molecule at different TE-TM splitting.} (a) Dependence of eigenfrequencies $\omega$ (in THz) of eigenmodes on the mode index $n$ for different TE-TM splittings: $\Delta_\text{LT}=$0 (orange dots), 0.2 (green dots), and 0.4 meV (blue dots). The inset shows the profile of the hexagonal ring potential. (b) Amplitude $|\Psi_{\pm}|$ and phase arg$(\Psi_{\pm})$ distributions, marked by the mode indices corresponding to the mode sequence in (a), for both polarization components $\Psi_{+}$ and $\Psi_{-}$ at $\Delta_\text{LT}=0$ meV (orange numbers) and $\Delta_\text{LT}=0.2$ meV (green numbers).}
\label{fig:EigenStates}
\end{figure*}

The evolution of two polarization components of polariton condensates under the coherent pump can be described by the spinor Gross-Pitaevskii equations~\cite{PhysRevLett.114.116401,kartashov2016modulational}:
\begin{align}\label{eq:model}
i\hbar\frac{\partial\Psi_{\pm}(\text{r},t)}{\partial t} &=\left[-\frac{\hbar^2}{2m}\triangledown_{\perp}^{2}-i\hbar\frac{\gamma_{\text{c}}}{2}+g_{\text{c}}|\Psi_{\pm}|^2+V(\textbf{r})\right]\Psi_{\pm}(\text{r},t) \nonumber \\ 
&+\frac{\Delta_\text{LT}}{k^2}\left(\partial_{x}\mp{i}\partial_{y}\right)^2\Psi_{\mp}(\text{r},t)+E_{\pm}(\text{r},t). 
\end{align}
Here, the indices $\pm$ indicate the right-/left-circular polarization components of polaritons, the effective polariton mass is given by $m=10^{-4}{m_{\text{e}}}$ (${m_{\text{e}}}$ is the free electron mass),  $\gamma_{\text{c}}=0.02$ ps$^{-1}$ is the polariton loss rate, $g_{\text{c}}=2~\mu \textrm{eV}~\mu \textrm{m}^2$ denotes the polariton-polariton interaction strength, $\Delta_\text{LT}$ represents the TE-TM splitting (leading to the SOI) intrinsically present in microcavities at the in-plane momentum $k=2$ $\mu$m$^{-1}$, $V(\textbf{r})=\sum_{\text{n=1}}^6{\mathcal{V}}(x-x_{\text{n}},y-y_{\text{n}})$ is the potential energy landscape created by micropillars with the diameter $2d=2$ $\mu$m arranged into a ring with a radius of $R=2$ $\mu$m so that neighboring pillars slightly touch each other [see the inset in Fig. \ref{fig:EigenStates}(a)], ${\mathcal{V}}=V_{0}e^{-(x^2+y^2)^5/d^{10}}$ describes the contribution from the individual pillar with depth $V_0=-5$ meV, and $E_{\pm}(\text{r},t)$ is the coherent pump.

We first analyze the linear eigenstates of the hexagonal molecule in the conservative regime, by setting $\gamma_\text{c}=0$ and $E_{\pm}=0$. Assuming linear solutions of the form $\Psi_{\pm}(\text{r},t)=u_{\pm}(\text{r})e^{-i\omega{t}}$ we obtain from \eqref{eq:model} the eigenvalue problem ${\hbar \omega}u_{\pm}=-\frac{\hbar^2}{2m}\triangledown_{\perp}^{2}u_{\pm}+V u_{\pm}+\frac{\Delta_\text{LT}}{k^2}\left(\partial_{x}\mp{i}\partial_{y}\right)^2u_{\mp}$
that we solved to obtain eigenfrequencies $\omega$ and shapes $u_{\pm}$ of the linear modes. If we consider only the lowest fundamental mode in each pillar (note that in a single micropillar the higher-order modes of the polariton condensates with integer or fractional orbital angular momenta can also be observed~\cite{PhysRevB.97.195149,sedov2020persistent}), the whole potential with six pillars give rise to 12 modes (Fig.~\ref{fig:EigenStates}). In the absence of the TE-TM splitting $\Delta_\text{LT}=0$ meV, the two polarization components are decoupled, the $\Psi_+$ and $\Psi_-$ distributions can, for example, have the same phase or $\pi$ phase difference for the same spatial distribution. Therefore there exist at least six sets of degenerated states. An additional degeneracy is connected with the fact that for a selected component $\Psi_+$ or $\Psi_-$ in the six-pillar structure only vortices with topological charges (winding numbers) of $|m|\leq 2$ are allowed and pairs of states with charges $+m$ and $-m$ are degenerate too (their superposition gives multipole states), with the exception for the state with $m=0$ that remains singlet, and for the singlet multipole state whose wavefunction changes its sign in each pillar that replaces the forbidden $m=3$ vortex~\cite{PhysRevLett.95.123902}. The modes shown in Fig.~\ref{fig:EigenStates} labelled with orange numbers are building blocks, whose linear combinations give $m=\pm1$ vortices (modes 3,4 and 5,6, $\omega=$-5.665 THz) and $m=\pm2$ vortices (modes 7,8 and 9,10, $\omega=$-5.286 THz). These two latter sets of modes form quadruplets and in each quadruplet only two pairs of modes have different density distributions. The modes with lowest and highest frequencies form doublets with identical density distributions.

\begin{figure*} [t]
\centering
{\includegraphics[width=1\linewidth]{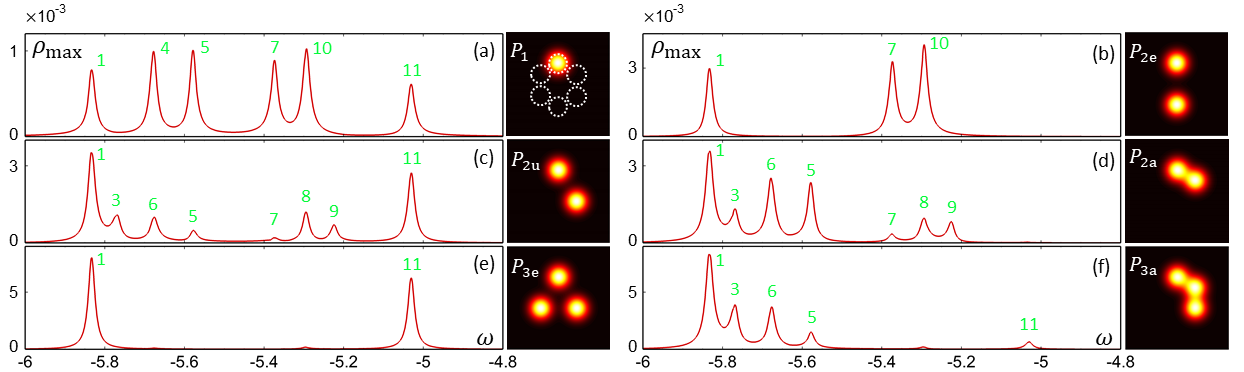}}
\caption{\textbf{Linear modes excited for different pump configurations.} Dependencies of the peak density ($\rho_\text{max}$ in $\mu$m$^{-2}$) of $\Psi_{+}$ on the pump frequency ($\omega$ in THz) for different pump configurations: (a) one-pillar excitation $P_{1}$, (b) two pump spots located symmetrically $P_\text{2e}$, (c) two pump spots asymmetrically located on even pillars $P_\text{2u}$, (d) two pump spots in adjacent pillars $P_\text{2a}$, (e) three symmetric pump spots $P_\text{3e}$, and (f) three pump spots in adjacent pillars $P_\text{3a}$. Pump profiles are shown to the right of the $\rho_{\textrm{max}}(\omega)$ dependencies. The potential landscape (dashed circles) is superimposed on top of the pump profile in (a). The green indices close to the peaks correspond to the modes marked by the green numbers in Fig.~\ref{fig:EigenStates}(b). Here, $E_{\pm}^{0}=$0.000001 and $\Delta_\text{LT}=0.2$ meV.}
\label{fig:LinearModes}
\end{figure*}

The presence of the TE-TM splitting breaks the symmetry of the system and lifts the degeneracy of the vortex modes with opposite topological charges, see the green and blue dots in Fig.~\ref{fig:EigenStates}(a) for different values of the TE-TM splitting. Remarkably, two lowest modes 1,2 and two highest modes 11,12 remain degenerate and experience a shift toward lower frequency values that progressively increases with the increase of the TE-TM splitting (compare green dots for $\Delta_\text{LT}=0.2$ meV with blue ones for $\Delta_\text{LT}=0.4$ meV). These modes are not vortices [see modes for green numbers 1,2 and 11,12 in Fig.~\ref{fig:EigenStates}(b)]. Two modes from each quadruplet remain degenerate and their frequencies decrease (modes 4,6 and 8,10 in Fig.~\ref{fig:EigenStates}(b)]. Another two modes from each quadruplet acquire strong and opposite frequency shifts and become non-degenerate (modes 3,5 and 7,9). All these modes with green numbers 3-10 in Fig.~\ref{fig:EigenStates}(b) carry nonzero topological charges that are opposite in two polarization components. The modes emerging after splitting of the lower quadruplet carry vortices with topological charges $m=\pm1$, while modes emerging from the upper quadruplet carry vortices with charges $m=\pm2$. Note that phase singularities in modes 8 and 10 split into two singularities due to the presence of the TE-TM splitting. If the splitting is significant, the shift of the modes in the frequency domain may become so strong that the order of the modes changes completely: for example, the blue point for the mode $n=3$ may be pushed below the eigenfrequency of the vortex-free states.

To study the excitation dynamics of the vortex modes, we consider the coherent pump $E_{\pm}(\text{r},t)=E_{\pm}(\text{r})e^{-i\omega{t}}$, where $\omega$ is the frequency of the pump, and search for the stationary solutions of \eqref{eq:model} by solving the time-independent equation:
\begin{align}\label{stationaryGPE}
\left[-\frac{\hbar^2}{2m}\triangledown_{\perp}^{2}-i\hbar\frac{\gamma_{\text{c}}}{2}+g_{\text{c}}|u_{\pm}|^2+V(\textbf{r})\right]u_{\pm} \nonumber \\
+\frac{\Delta_\text{LT}}{k^2}\left(\partial_{x}\mp{i}\partial_{y}\right)^2u_{\mp}+E_{\pm}(\text{r})-\hbar\omega{u_{\pm}}=0. 
\end{align}
For the homogeneous pump, only the fundamental mode [green number 1 in Fig.~\ref{fig:EigenStates}(b)] can be excited. To excite the vortex modes, we use a Gaussian pump beam given by $E_{\pm}(\text{r})=E_{\pm}^{0}e^{-\textbf{r}^2/d^2}$, where $E_{\pm}^{0}$ is the amplitude of the pump and pump beam width $d=1$ $\mu$m is similar to that of the pillar [see Fig.~\ref{fig:LinearModes}(a)].

For a single Gaussian pump beam located at an arbitrary (for example, the upper) pillar, several non-degenerate modes can be excited by changing pump frequency $\omega$ at $\Delta_\text{LT}=0.2$ meV [Fig.~\ref{fig:LinearModes}(a)]. The excitation efficiency of the modes depends not only on the frequency of the pump, but also on projection of the pump on the linear mode profile. For selected pump configuration, this projection is maximal for modes $n=1, 4, 5, 7, 10, 11$ from Fig.~\ref{fig:EigenStates}(b), leading to resonant spikes 
at corresponding eigenfrequencies, as shown in Fig.~\ref{fig:LinearModes}(a). The variety of excited modes can be efficiently controlled by using more than one pump spot. In all cases the excitation efficiency is given by the magnitude of pump projection on the mode. For two symmetrically located pump spots [Fig.~\ref{fig:LinearModes}(b)], the phase of the condensate in these two pillars should be the same to have maximal projection for a given polarization component, which leads to the excitation of the modes $n=$1, 7, and 10. If we keep one pump spot in the upper pillar and move the second spot from the bottom pillar to its right neighbor [Fig.~\ref{fig:LinearModes}(c)], the two vortex-free modes 1 and 11 are strongly enhanced. For adjacent pump spots in Fig.~\ref{fig:LinearModes}(d) the phase distribution of the mode 11 does not match the pump distribution anymore and disappears, while vortex-carrying modes 5,6 with $m=\pm1$ are enhanced. Comparing the results in Fig.~\ref{fig:LinearModes}(a)-(d), one can see that the second pump spot does lead to resonances with  the vortex modes 3 and 9 that are absent for one-spot excitation. If three symmetrically located pump spots are used, as shown in Fig.~\ref{fig:LinearModes}(e), only non-vortex modes survive because projection of such pump on vortex-carrying states is nearly zero. The excitation of vortices requires asymmetric pump configurations, such as pump spots on three adjacent pillars, see Fig.~\ref{fig:LinearModes}(f). Three adjacent pump spots prevent the quick change of the phase of the condensates in the neighboring pillars. As a result, the higher-order vortex modes with $m=\pm2$ vanish. If pump is provided in each pillar only the fundamental mode will be excited, similarly to the case of the homogeneous excitation. The results of Fig.~\ref{fig:LinearModes} confirm that required states can always be excited by changing pump configuration even in the case, when pump itself does not carry a topological charge.

\begin{figure} [!h]
\centering
{\includegraphics[width=1\linewidth]{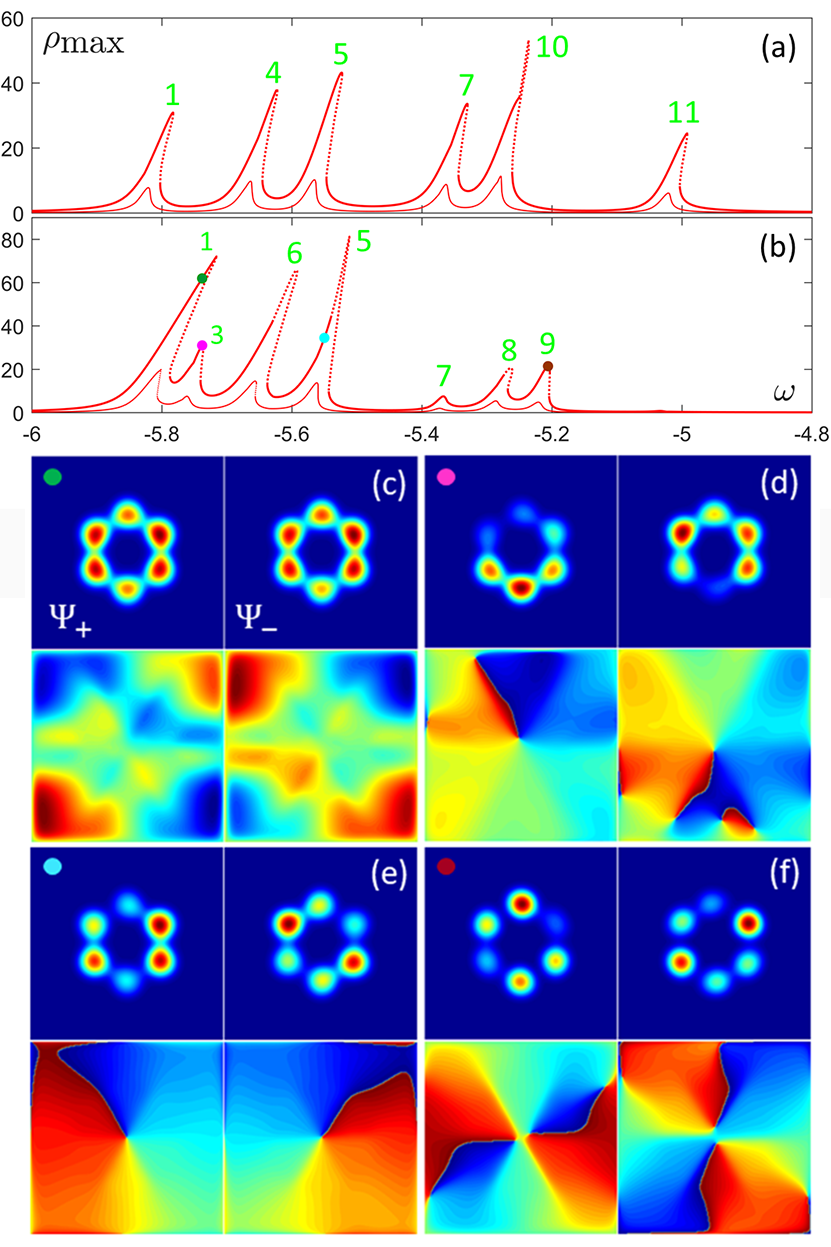}}
\caption{\textbf{Nonlinear modes.} Peak density ($\rho_\text{max}$ in $\mu$m$^{-2}$) of $\Psi_{+}$ vs pump frequency ($\omega$ in THz) for (a) one pump spot $P_{1}$ with $E_{\pm}^{0}=$0.0001 (thin line) and 0.0002 (thick line), and (b) two adjacent pump spots $P_\text{2a}$ with $E_{\pm}^{0}=$0.000075 (thin line) and 0.00015 (thick line). The modes corresponding to the peaks are marked by the mode indices, associated to the modes marked by the green numbers in Fig.~\ref{fig:EigenStates}(b). The solid (dotted) lines are the stable (unstable) solutions. (c-f) Density (top row) and phase (bottom row) profiles of the stable states, corresponding to the points in (b). Here, $\Delta_\text{LT}=0.2$ meV.}
\label{fig:NonlinearModes}
\end{figure}

For stronger pump intensities nonlinear effects become significant, resulting in blue shifts of the frequencies of the solutions. The resonance peaks in  $\rho_{\textrm{max}}(\omega)$ dependencies become tilted and broaden with the increase of pump amplitude [Fig.~\ref{fig:NonlinearModes}(a)]. In Fig.~\ref{fig:LinearModes}(d) one can see that the first two peaks at the left side are very close to each other, so that they are tilted then the nonlinearity versions start overlapping [see Fig.~\ref{fig:NonlinearModes}(b)] because the first stronger peak tilts more than the second weaker peak as the pump intensity increases. This leads to a bistability: the coexistence of two stable nonlinear states shown in Fig.~\ref{fig:NonlinearModes}(c,d) at the same value of $\omega$. Similar phenomena have been demonstrated in the similar potential landscapes for photon-like modes~\cite{zambon2019orbital} or under non-resonant excitation~\cite{barkhausen2020multistable}. Stable nonlinear states originating from the non-degenerate modes 5 and 9 can be found in Fig.~\ref{fig:NonlinearModes}(e,f). These solutions become strongly asymmetric. Increasing pump strength leads to the splitting of the vortex with a higher topological charge $m=\pm$2 [cf. the vortex mode 9 in Fig.~\ref{fig:EigenStates}(b) into two $m=\pm$1 ones as shown in Fig.~\ref{fig:NonlinearModes}(f) because of the nonlinearity-induced mixture of the modes.

To conclude, we have studied the formation of vortex polariton states in a hexagonal molecule composed by six micropillars. The vortices are generated under resonant excitation by Gaussian pulses due to the SOI. The vortex modes can be excited selectively by properly choosing the pump frequency and position. The nonlinearity leads to the blue-shift of the frequencies of the modes, resulting in the emergence of a bistability. Our findings pave the way to the realization of vortex polariton lasers where lasing from topologically protected modes would be realised.

\textbf{Funding.} Deutsche Forschungsgemeinschaft (DFG) (No. 231447078, 270619725); Russian Science Foundation (Project 17-12-01413-$\Pi$). Westlake University, project 041020100118 and the Program 2018R01002 funded by Leading Innovative and Entrepreneur Team Introduction Program of Zhejiang.

\textbf{Acknowledgements.} Paderborn Center for Parallel Computing, PC$^2$; A.K. acknowledges Saint-Petersburg State University support program, ID 40847559.

\textbf{Disclosures.} The authors declare no conflicts of interest.


\begin{thebibliography}{10}
\newcommand{\enquote}[1]{``#1''}

\bibitem{deng2002condensation}
H.~Deng, G.~Weihs, C.~Santori, J.~Bloch, and Y.~Yamamoto,
  {\protect\JournalTitle{Science}} \textbf{298}, 199 (2002).

\bibitem{kasprzak2006bose}
J.~Kasprzak, M.~Richard, S.~Kundermann, A.~Baas, P.~Jeambrun, J.~M.~J. Keeling,
  F.~M. Marchetti, M.~H. Szyma{\'n}ska, R.~Andr{\'e}, J.~L. Staehli, V.~Savona,
  P.~B. Littlewood, B.~Deveaud, and L.~S. Dang, {\protect\JournalTitle{Nature}}
  \textbf{443}, 409 (2006).

\bibitem{PhysRevA.53.4250}
A.~Imamoglu, R.~J. Ram, S.~Pau, and Y.~Yamamoto, {\protect\JournalTitle{Phys.
  Rev. A}} \textbf{53}, 4250 (1996).

\bibitem{wolf2001spintronics}
S.~A. Wolf, D.~D. Awschalom, R.~A. Buhrman, J.~M. Daughton, S.~von Moln{\'a}r,
  M.~L. Roukes, A.~Y. Chtchelkanova, and D.~M. Treger,
  {\protect\JournalTitle{Science}} \textbf{294}, 1488 (2001).

\bibitem{bliokh2015spin}
K.~Y. Bliokh, F.~J. Rodr{\'\i}guez-Fortu{\~n}o, F.~Nori, and A.~V. Zayats,
  {\protect\JournalTitle{Nature Photonics}} \textbf{9}, 796 (2015).

\bibitem{zambon2019optically}
N.~C. Zambon, P.~St-Jean, M.~Mili{\'c}evi{\'c}, A.~Lema{\^\i}tre, A.~Harouri,
  L.~Le~Gratiet, O.~Bleu, D.~D. Solnyshkov, G.~Malpuech, I.~Sagnes, S.~Ravets,
  A.~Amo, and J.~Bloch, {\protect\JournalTitle{Nature Photonics}} \textbf{13},
  283 (2019).

\bibitem{PhysRevB.59.5082}
G.~Panzarini, L.~C. Andreani, A.~Armitage, D.~Baxter, M.~S. Skolnick, V.~N.
  Astratov, J.~S. Roberts, A.~V. Kavokin, M.~R. Vladimirova, and M.~A.
  Kaliteevski, {\protect\JournalTitle{Phys. Rev. B}} \textbf{59}, 5082 (1999).

\bibitem{PhysRevLett.92.017401}
K.~V. Kavokin, I.~A. Shelykh, A.~V. Kavokin, G.~Malpuech, and P.~Bigenwald,
  {\protect\JournalTitle{Phys. Rev. Lett.}} \textbf{92}, 017401 (2004).

\bibitem{PhysRevX.5.011034}
V.~G. Sala, D.~D. Solnyshkov, I.~Carusotto, T.~Jacqmin, A.~Lema\^{\i}tre,
  H.~Ter\ifmmode~\mbox{\c{c}}\else \c{c}\fi{}as, A.~Nalitov, M.~Abbarchi,
  E.~Galopin, I.~Sagnes, J.~Bloch, G.~Malpuech, and A.~Amo,
  {\protect\JournalTitle{Phys. Rev. X}} \textbf{5}, 011034 (2015).

\bibitem{PhysRevLett.99.106401}
Y.~G. Rubo, {\protect\JournalTitle{Phys. Rev. Lett.}} \textbf{99}, 106401
  (2007).

\bibitem{lagoudakis2009observation}
K.~G. Lagoudakis, T.~Ostatnick{\`y}, A.~V. Kavokin, Y.~G. Rubo, R.~Andr{\'e},
  and B.~Deveaud-Pl{\'e}dran, {\protect\JournalTitle{Science}} \textbf{326},
  974 (2009).

\bibitem{PhysRevB.94.134310}
A.~V. Yulin, A.~S. Desyatnikov, and E.~A. Ostrovskaya,
  {\protect\JournalTitle{Phys. Rev. B}} \textbf{94}, 134310 (2016).

\bibitem{PhysRevB.101.205301}
M.~Pukrop, S.~Schumacher, and X.~Ma, {\protect\JournalTitle{Phys. Rev. B}}
  \textbf{101}, 205301 (2020).

\bibitem{PhysRevLett.95.136601}
A.~Kavokin, G.~Malpuech, and M.~Glazov, {\protect\JournalTitle{Phys. Rev.
  Lett.}} \textbf{95}, 136601 (2005).

\bibitem{leyder2007observation}
C.~Leyder, M.~Romanelli, J.~P. Karr, E.~Giacobino, T.~C. Liew, M.~M. Glazov,
  A.~V. Kavokin, G.~Malpuech, and A.~Bramati, {\protect\JournalTitle{Nature
  Physics}} \textbf{3}, 628 (2007).

\bibitem{PhysRevLett.110.016404}
H.~Flayac, D.~D. Solnyshkov, I.~A. Shelykh, and G.~Malpuech,
  {\protect\JournalTitle{Phys. Rev. Lett.}} \textbf{110}, 016404 (2013).

\bibitem{donati2016twist}
S.~Donati, L.~Dominici, G.~Dagvadorj, D.~Ballarini, M.~De~Giorgi, A.~Bramati,
  G.~Gigli, Y.~G. Rubo, M.~H. Szyma{\'n}ska, and D.~Sanvitto,
  {\protect\JournalTitle{Proceedings of the National Academy of Sciences}}
  \textbf{113}, 14926 (2016).

\bibitem{PhysRevLett.114.116401}
A.~V. Nalitov, D.~D. Solnyshkov, and G.~Malpuech, {\protect\JournalTitle{Phys.
  Rev. Lett.}} \textbf{114}, 116401 (2015).

\bibitem{PhysRevB.91.161413}
C.-E. Bardyn, T.~Karzig, G.~Refael, and T.~C.~H. Liew,
  {\protect\JournalTitle{Phys. Rev. B}} \textbf{91}, 161413 (2015).

\bibitem{PhysRevX.5.031001}
T.~Karzig, C.-E. Bardyn, N.~H. Lindner, and G.~Refael,
  {\protect\JournalTitle{Phys. Rev. X}} \textbf{5}, 031001 (2015).

\bibitem{kartashov2016modulational}
Y.~V. Kartashov and D.~V. Skryabin, {\protect\JournalTitle{Optica}} \textbf{3},
  1228 (2016).

\bibitem{klembt2018exciton}
S.~Klembt, T.~H. Harder, O.~A. Egorov, K.~Winkler, R.~Ge, M.~A. Bandres,
  M.~Emmerling, L.~Worschech, T.~C.~H. Liew, M.~Segev, C.~Schneider, and
  S.~H\"{o}fling, {\protect\JournalTitle{Nature}} \textbf{562}, 552 (2018).

\bibitem{PhysRevLett.122.083902}
Y.~V. Kartashov and D.~V. Skryabin, {\protect\JournalTitle{Phys. Rev. Lett.}}
  \textbf{122}, 083902 (2019).

\bibitem{zhang2019finite}
W.~Zhang, X.~Chen, Y.~V. Kartashov, D.~V. Skryabin, and F.~Ye,
  {\protect\JournalTitle{Laser \& Photonics Reviews}} \textbf{13}, 1900198
  (2019).

\bibitem{PhysRevB.75.241301}
T.~C.~H. Liew, A.~V. Kavokin, and I.~A. Shelykh, {\protect\JournalTitle{Phys.
  Rev. B}} \textbf{75}, 241301 (2007).

\bibitem{ma2020realization}
X.~Ma, B.~Berger, M.~A{\ss}mann, R.~Driben, T.~Meier, C.~Schneider,
  S.~H{\"o}fling, and S.~Schumacher, {\protect\JournalTitle{Nature
  Communications}} \textbf{11}, 1 (2020).

\bibitem{PhysRevB.97.195149}
V.~A. Lukoshkin, V.~K. Kalevich, M.~M. Afanasiev, K.~V. Kavokin,
  Z.~Hatzopoulos, P.~G. Savvidis, E.~S. Sedov, and A.~V. Kavokin,
  {\protect\JournalTitle{Phys. Rev. B}} \textbf{97}, 195149 (2018).

\bibitem{sedov2020persistent}
E.~Sedov, V.~Lukoshkin, V.~Kalevich, Z.~Hatzopoulos, P.~Savvidis, and
  A.~Kavokin, {\protect\JournalTitle{ACS Photonics}} \textbf{7}, 1163 (2020).

\bibitem{PhysRevLett.95.123902}
Y.~V. Kartashov, A.~Ferrando, A.~A. Egorov, and L.~Torner,
  {\protect\JournalTitle{Phys. Rev. Lett.}} \textbf{95}, 123902 (2005).

\bibitem{zambon2019orbital}
N.~C. Zambon, P.~St-Jean, A.~Lema{\^\i}tre, A.~Harouri, L.~Le~Gratiet,
  I.~Sagnes, S.~Ravets, A.~Amo, and J.~Bloch, {\protect\JournalTitle{Optics
  letters}} \textbf{44}, 4531 (2019).

\bibitem{barkhausen2020multistable}
F.~Barkhausen, S.~Schumacher, and X.~Ma, {\protect\JournalTitle{Optics
  Letters}} \textbf{45}, 1192 (2020).

\end{thebibliography}
\end{document}